\documentclass[pre,twocolumn,showpacs,showkeys,preprintnumbers,amsmath,amssymb]{revtex4}
\usepackage{graphicx}
\usepackage{dcolumn}
\usepackage{bm}
\usepackage{subfigure}
\usepackage{amssymb}
\usepackage{graphics}
\usepackage{epsfig}
\usepackage{threeparttable}
\begin{document}
\title{Nonequilibrium dynamics of a  spin-3/2 Blume Capel model with quenched random crystal field}
\author{Erol Vatansever}
\author{Hamza Polat}\email{hamza.polat@deu.edu.tr}
\affiliation{Department of Physics, Dokuz Eyl\"{u}l University, TR-35160 Izmir, Turkey}
\date{\today}
\begin{abstract}
The relaxation and complex magnetic susceptibility treatments of a spin-3/2 Blume-Capel model with quenched random
crystal field on a two dimensional square lattice are investigated by a method combining the statistical
equilibrium theory and the thermodynamics of linear irreversible processes. Generalized force and flux are
defined in irreversible thermodynamics limit. The kinetic  equation for  the magnetization is obtained
by using linear response theory. Temperature and also crystal field dependencies of the relaxation time
are obtained in the vicinity of phase transition points. We found that the relaxation time exhibits divergent
treatment near the order-disorder phase transition point as well as near the isolated critical point
whereas it displays cusp behavior near the first order phase transition point. In addition, much effort has been devoted
to  investigation of complex magnetic susceptibility response of the system to changing applied field frequencies
and it is observed that the considered disordered magnetic system exhibits unusual and interesting behaviors.
Furthermore, dynamical mean field critical exponents for the relaxation time and complex magnetic susceptibility
are calculated in order to formulate the critical behavior of the system.
Finally, a comparison of our observations with those of recently published studies is represented and
it is shown that there exists a qualitatively good agreement.
\end{abstract}
\maketitle
\section{Introduction}\label{intro}
Investigation of disorder effects on the critical phenomena has a long history and there have been a great  many of
theoretical studies focused on disordered magnetic materials with quenched randomness where the random variables of
a magnetic system such as random fields \cite{larkin, imry} or random bonds \cite{edwards, sherington} may not
change its value over time. On the other hand, quenched crystal field diluted ferromagnets constitute another
example of magnetic systems and equilibrium properties of this type models have been investigated under
multifarious approximation methods \cite{benyoussef1, branco, kaneyoshi2, boccara2}. Besides, effects of
disorder on magnetic systems have been systematically studied, not only for theoretical interests
but also for the identifications with experimental realizations \cite{bouchiat, katsumata, belanger}.
It has been shown by renormalization group  arguments that first-order transitions are replaced by
continuous transitions, consequently tricritical points and critical end points are depressed in
temperature, and a finite amount of disorder will suppress them \cite{huifaliong}.

Since the time in which Ising model \cite{Ising} was invented there exists a limited number of studies
including the dynamic nature of the system under the influence a small perturbation.
Even though the physical investigations regarding these systems bring about
a lot of mathematical difficulties, the nonequilibrium systems are in the focus of scientists
because they have an unusual and interesting dynamic behavior. It is known that
examination of physical properties of  the nonequilibrium systems allows us to analyze
various dynamical concepts. One of them is the  relaxation time of a considered system which exhibits a
divergence near the critical and  multicritical points and gives remarkable information about
dynamic properties. As far as we know,  on the theoretical picture, the relaxation behavior of a
limited number of systems has been investigated by making use of Onsager reciprocity theorem \cite{onsager1, onsager2}.
This type of investigation  was first performed by  Tanaka \emph{et al.} \cite{tanaka} near the
continuous  phase transition point for the spin-1/2 Ising model  by means of Onsager  reciprocity theorem.
Next, AB-type ferromagnetic and antiferromagnetic models have been  investigated by Barry \cite{barry1},
Barry and Harrington \cite{barry2} using the similar formulation, respectively.
Additionally, in Ref. \cite{barry2}, dynamic initial parallel susceptibility expression is analyzed  near
the N\'{e}el temperature  for considered applied field frequencies. Recently, spin-1 Ising model has been analyzed
by Erdem and Keskin within the framework of the entropy
production \cite{erdem1, erdem2} and it is found that one of the relaxation times diverges at the  continuous
phase transition point while the other relaxation time remains finite. The relaxation process of  the pure spin-3/2
Blume-Emery-Griffiths (BEG) model with bilinear and biquadratic exchange interactions
has been analyzed near the critical \cite{keskin} and multicritical points \cite{canko}.  In addition, thermal
variations of the relaxation time of a metamagnetic Ising model has been investigated by Gulpinar \emph{et al.} \cite{gulpinar1}.
Very recently, following the same methodology, spin-1 Blume-Capel (BC) model with quenched random crystal field has been studied
in detail and it is observed that the relaxation time has a jump discontinuity at the first-order phase transition point
whereas it diverges in the neighborhood of multicritical points such as tricritical,
critical end point and bicritical end points \cite{gulpinar2}. Moreover,
from the experimental point of view, a great many of experimental studies have been dedicated  to the better understanding of
dynamic critical phenomena in magnetic systems \cite{han, eesley, collins, schuller, reitze}.

On the other side, it is known that the dynamic or ac susceptibility measurement, in which a time varying oscillating magnetic
field is carried out to a sample, is a powerful method for analyzing the dynamic evolution of the considered real magnetic
system. It is obtained from the dynamic response of the system to time dependent magnetic field and  many studies
have been performed  regarding the magnetic relaxation of cooperatively interacting systems, such as nanoparticles
\cite{raap, bhowmik, sharma}, spin glasses \cite{kotzler}, high $T_{c}$ systems \cite{engelstad} and magnetic fluids \cite{fannin}.
From the theoretical point of view, in Ref. \cite{acharyya}, Acharyya and Chakrabarti probed thoroughly the real and imaginary
parts of magnetic susceptibility in the neighborhood of phase transition temperature of a spin-1/2 Ising model under the
time dependent oscillating external magnetic field by using Monte Carlo (MC) simulation with periodic boundary conditions.
Additionaly, Erdem investigated the magnetic relaxation in a spin-$1$ Ising model near the second-order phase
transition point where time derivatives of the dipolar and quadrupolar order parameters are treated as fluxes
conjugate to their appropriate generalized forces in the sense of irreversible thermodynamics \cite{erdem3}, next,
the same author has analyzed the frequency dependence of the complex susceptibility for the same system \cite{erdem4}.
In addition to these, with the helping of Nelson's method, a systematic investigation containing frequency, momentum
and temperature dependent response function has been achieved for Ising-type system below the
critical temperature \cite{pawlak}. Very recently, a comprehensive study including the
equilibrium and nonequilibrium antiferromagnetic and ferromagnetic susceptibilities
of a metamagnetic Ising system  in the vicinity of order-disorder
transition point has been done by benefiting from mean field approximation (MFA) \cite{gulpinar3}.

Apart from these, the effects of impurities on driving-rate-dependent energy loss in a
ferromagnet under the time dependent magnetic field have been analyzed by Zheng and Li \cite{zheng_li}
by using several well defined models  within the frameworks of MFA and MC, and they found using MFA that,
the hysteresis loop area is a power law function of the linear driving rate as $A-A_{0}\propto h^{\beta}$,
where, $A_{0}, h$ and  $\beta$ are the static hysteresis loop area, the linear driving rate and scaling
exponent of the system, respectively. Very recently,  the quenched site and bond diluted kinetic Ising models
under the influence of a time dependent oscillating magnetic field have been analyzed by making use of
effective field theory \cite{akinci, vatansever} on a two dimensional honeycomb
lattice. In Ref. \cite{akinci},  the global phase diagrams including the reentrant phase transitions are presented
by the authors for site diluted kinetic Ising model, and they  showed that the coexistence regions
disappear for sufficiently weak dilution of lattice sites. Following the same methodology, the authors have
concentrated on the influences of quenched bond dilution process on the dynamic behavior of the system.
After some detailed analysis, it has been found that the impurities in a bond diluted kinetic Ising model give
rise to a number of interesting  and unusual phenomena such as reentrant phenomena and the impurities to
have a tendency to destruct the first-order transitions and the dynamic tricritical point \cite{vatansever}.
Furthermore,  it has also been shown that dynamically ordered phase regions get expanded with decreasing
amplitude which is more evident at low frequencies.

As far as we know, the nonequilibrium properties of the quenched disordered systems such
as crystal field (and also bond or site) diluted ferromagnets under the influence of a small magnetic field
perturbation have not yet been investigated  for higher spin models.
These types of disorder effects constitute an important role in the real magnetic material science,
since the quenched disorder effects may induce some important macroscopic influences on the material.
Therefore, we believe that the investigation of the effects of quenched randomness on the nonequilibrium
properties of the Ising model and its derivations still need  particular attention. Hence, in this work based on MFA,
we intend to study the nonequilibrium properties of a spin-3/2 BC model by introducing the quenched crystal
field dilution effects.  Here, we used a  method combining statistical equilibrium theory and the thermodynamics of
irreversible processes to analyze the magnetic relaxation behavior and dynamic complex susceptibility of a
spin-3/2 BC model with quenched crystal field randomness. In this context, by means of the
MFA to obtain magnetic Gibbs free-energy, a generalized force and a generalized current are
determined  within the framework of irreversible thermodynamics in the neighborhood of equilibrium states.
After that, the kinetic equation for the time dependent magnetization is obtained and a relaxation time
is derived and also thermal and crystal field variations are
examined in the neighborhood of critical and isolated critical points. In addition, we calculated the
complex magnetic susceptibility expression and analyzed near the isolated critical as well as second order
phase transition points for selected Hamiltonian parameters. It can be said that the considered
disordered magnetic system  exhibits unusual and interesting behaviors. The aforementioned features will be
discussed in detail in later sections.

The remaining part  of the paper is as follows: In Sec. \ref{formulation} we briefly present the formulations,
the results and discussions are presented in Sec. \ref{results}, and finally Sec. \ref{conclude} contains our conclusions.

\section{Formulation}\label{formulation}
In this section, we give the formulation of present study for a two-dimensional square lattice.
The Hamiltonian of the spin-3/2 BC model is given by:
\begin{equation}\label{eq1}
\hat{H}=-J\sum_{\langle ij\rangle}S_{i}S_{j}+\sum_{i}\Delta_{i}S_{i}^{2},
\end{equation}
where the first term is a summation over the nearest neighbor spins with $S_{i}=\pm3/2, \pm1/2$ and $\Delta_{i}$,
in the second term, represents random crystal field which is distributed according to a given probability
distribution function:
\begin{equation}\label{eq2}
G(\Delta_{i})=\frac{1}{2}\left(\delta(\Delta_{i}-\Delta(1+\alpha))+\delta(\Delta_{i}-\Delta(1-\alpha)) \right),
\end{equation}
here $\alpha\geq0$, and half of the lattice sites subject to a crystal  field $\Delta(1+\alpha)$ and the
remaining lattice sites have a crystal field $\Delta(1-\alpha)$. In order to examine the nonequilibrium properties
of the considered magnetic system, first we have to focus on the equilibrium properties following the
definition given in Ref. \cite{bahmad}. In this context, we have derived the Gibbs free energy  $(G=U-TS-BNm)$ by making
use of MFA in the following form:
\begin{equation}\label{eq3}
\begin{array}{lcl}
&&g(m, t, d, b)=G/N=\frac{zm^{2}}{2}-bm \\
\\
&&-\frac{t}{2}\left[\log\left(2\exp\left(-\frac{9\ell_{1}}{4}\right)\ell_{3}+2\exp\left(-\frac{\ell_{1}}{4}\right)\ell_{4}\right)\right]\\
\\
&&-\frac{t}{2}\left[\log\left(2\exp\left(-\frac{9\ell_{2}}{4}\right)\ell_{3}+
2\exp\left(-\frac{-\ell_{2}}{4}\right)\ell_{4}\right)\right].
\end{array}
\end{equation}
where $z$, $t=T/J$, $b=B/J$ and $d=\Delta/J$ are the coordination number, the reduced temperature,
the reduced magnetic field and reduced crystal field, respectively. In addition, for the sake of simplicity,
the coefficients $\ell_{i} (i=1,..,4)$ are defined as follows:
\begin{equation}\label{eq4}
\begin{array}{cc}
  \;\;\ell_{1}=\frac{d(1+\alpha)}{t}, & \; \; \; \; \;\; \;\ell_{3}=\cosh\left(\frac{3zm}{2t}\right), \\
   \\
  \; \;\ell_{2}=\frac{d(1-\alpha)}{t}, & \; \; \; \; \; \; \ell_{4}=\cosh\left(\frac{zm}{2t}\right). \\
  \\
\end{array}
\end{equation}
In the absence of the external applied field $(b)$ and by benefiting  from the minimization condition of the free energy
at the equilibrium $\left(\frac{\partial g}{\partial m}=0\right)$, the mean field equation of state is found as follows:
\begin{equation}\label{eq5}
m=\frac{3\ell_{5}+e^{2\ell_{1}}\ell_{6}}{4\ell_{3}+4e^{2\ell_{1}}\ell_{4}}+\frac{3\ell_{5}+
e^{2\ell_{2}}\ell_{6}}{4\ell_{3}+4e^{\ell_{2}}\ell_{4}}.
\end{equation}
here $\ell_{5}=\sinh\left(\frac{3zm}{2t}\right)$ and $\ell_{6}=\sinh\left(\frac{zm}{2t}\right)$. This equation may
be solved without difficultly by an iterative procedure, i.e. Newton-Raphson method.  Because the detailed equilibrium
investigations of  the considered system  are analyzed in Ref. \cite{bahmad}, we shall only give a brief
summary here as follows: Equilibrium phase diagrams of the spin-3/2 BC model sensitively depend on the selected value
of $\alpha$, and the system exhibits four kinds of behavior: For $0\leq \alpha <1/4$, phase transition line between
ferromagnetic  $F_{3/2}$ and the ferromagnetic  $F_{1/2}$ phases consists of only first order phase transition points
and also it is found that $\alpha=1/4$ is independent of the coordination number. Two successive
discontinuous phase transition lines between $F_{3/2}$ and $F_{1}$ phases  and between  $F_{1}$ and $F_{1/2}$ phases
occur for  the interval $1/4\leq \alpha <1$. For $\alpha=1$, a discontinuous phase transition line exists
separating $F_{3/2}$ and $F_{1}$ phases. On the other hand,
two discontinuous phase transition lines emerge for $\alpha>1$, the aforementioned transitions are
between $F_{1}$ and $F_{3/2}$ phases, and between  $F_{3/2}$ and $F_{1}$ phases.

\subsection{Magnetic Gibbs energy production, kinetic equations and relaxation time}
In order to understand the relaxation behavior of the spin-3/2 BC model with quenched
random crystal field,  one supposes a small uniform  external magnetic
field is applied  along $z$ axis.  When a system is disturbed from its equilibrium state, quantities  decay rapidly to
their equilibrium values \cite{reichl}. In that case,  new energy form of the considered magnetic system in
the neighborhood of equilibrium state can be  expressed in the following:
\begin{equation}\label{eq6}
g(m,t,d,b)=g^{0}(m_{0},t,d,b_{0})+\Delta g,
\end{equation}
where  $g^{0}(m_{0},t,d,b_{0})$ is equilibrium magnetic Gibbs energy in the absence of $b$, and $\Delta g$ is the energy
variation of the system and it is expressed as follows:
\begin{equation}\label{eq7}
\begin{array}{lcl}
\Delta g&=&\frac{1}{2}\Lambda_{1}(m-m_{0})^{2}+\Lambda_{2}(b-{b_{0}})\\
\\
 &&+\frac{1}{2}\Lambda_{3}(b-b_{{0}})^{2}+\Lambda_{4}(m-m_{0})(b-{b_{0}}),
\end{array}
\end{equation}
where the coefficients $\Lambda_{i} (i=1,..,4)$ can be easily calculated by using Taylor expansion
and  are given in Appendix.
According to theory of  irreversible thermodynamics, the generalized force $(X_{m})$, which conjugates the
current $(\dot{m})$, is derived by differentiating $\Delta g$ with respect to $m-m_{0}$:
\begin{equation}\label{eq8}
\displaystyle X_{m}=\displaystyle \frac{\partial \Delta g}{\partial (m-m_{0})}=\Lambda_{1}(m-m_{0})+\Lambda_{4}(b-b_{0}).
\end{equation}
The magnetic system will be relaxed back to its equilibrium state with a small deviation in the  external magnetic field strength,
and related process can be  investigated by using linear response theory because the deviation of the applied external magnetic
field is so small. The linear relation between the generalized flux and force can be written in terms of phenomenological coefficient as:
\begin{equation}\label{eq9}
\dot{m}=\gamma X_{m},
\end{equation}
\begin{equation}\label{eq10}
\dot{m}=\gamma \Lambda_{1}(m-m_{0})+\gamma \Lambda_{4}(b-b_{0}),
\end{equation}
where $\gamma$ is a kinetic rate coefficient. By assuming a simple form for the solution
$m-m_{0}\simeq \exp(-t/\tau)$, the linearized equation of motion has been solved and
the relaxation time can be found in the following form:
\begin{equation}\label{eq11}
\tau=-\frac{1}{\gamma\Lambda_{1}},
\end{equation}
By solving Eq. (\ref{eq11}) numerically for a combination of Hamiltonian parameters  we obtained the
dynamic evolution of the relaxation time of the spin-3/2 BC model with quenched random crystal field.

\subsection{Derivation of kinetic equations leading to complex magnetic susceptibility}
When a ferromagnetic material is subject to a periodically varying time-dependent magnetic field at an angular
frequency $\omega$, all quantities will oscillate near the equilibrium state at this same angular frequency:

\begin{equation}\label{eq12}
m-m_{0}=m_{1}\exp{(i\omega t)},   \quad    b-b_{0}=b_{1}\exp{(i\omega t)}.
\end{equation}
One can easily obtain the following form substituting Eq. (\ref{eq12}) into the kinetic equation
given by Eq. (\ref{eq10}):
\begin{equation}\label{eq13}
m_{1}\left(\gamma \Lambda_{1}-i\omega\right)+\gamma \Lambda_{4}b_{1}=0.
\end{equation}
Solving Eq. (\ref{eq13}) for $m_{1}/b_{1}$ gives:
\begin{equation}\label{eq14}
\frac{m_{1}}{b_{1}}=\frac{\gamma\Lambda_{4}}{i\omega -\gamma \Lambda_{1}}.
\end{equation}

Eq. (\ref{eq14}) can be used to  calculate the complex magnetic susceptibility of the considered system $\chi(\omega)$ and
the spin-3/2 BC model induced magnetization (total induced magnetic moment per unit volume) is given by:
\begin{equation}\label{eq15}
m-m_{\infty}=\mathrm{Re}\left(m_{1}\exp(i\omega t)\right),
\end{equation}
where $m_{\infty}$ is the magnetization induced by a time dependent oscillating magnetic field.  Also, by benefiting from definition,
the expression for  $\chi(\omega)$ may be written:
\begin{equation}\label{eq16}
m-m_{\infty}=\mathrm{Re}\left(\chi(\omega)b_{1}\exp(i\omega t)\right),
\end{equation}
where  $\chi(\omega)=\chi^{'}(\omega)-i\chi^{''}(\omega)$ is the complex magnetic susceptibility and  its real and imaginary parts are
called as magnetic dispersion and absorption factors, respectively. Finally the magnetic  dispersion and  absorption factors
can be found in the following forms:
\begin{equation}\label{eq17}
\begin{array}{lll}
\displaystyle\chi^{'}(\omega)&=&\displaystyle\frac{\gamma\Lambda_{4}\tau}{1+\omega^{2}\tau^{2}},\\
\\
\displaystyle\chi^{''}(\omega)&=&\displaystyle\frac{\gamma\Lambda_{4}\tau^{2}\omega}{1+\omega^{2}\tau^{2}}.\\
\end{array}
\end{equation}

\section{Results and Discussion}\label{results}
In this section, we will focus our attention on the nonequilibrium dynamics properties of the spin-3/2 BC model
with quenched crystal field. This section is divided into two parts as follows: In section \ref{results1}, we
discuss  how the quenched random crystal field affect the relaxation time behavior in the vicinity of
order-disorder  transition temperature as well as near the isolated critical point. In addition, we
examine the crystal  field variation of the relaxation time for selected Hamiltonian parameters,
and also we calculate dynamical mean field critical exponents in order to formulate the critical behavior
of relaxation time of the considered system. Real and imaginary parts of the complex magnetic susceptibility are
analyzed near the second-order phase transition temperature and also near the isolated critical point in
section \ref{results2}. As a final investigation, dynamical mean field critical exponents are
calculated for magnetic dispersion and absorption factors for both low and high frequency regions.

\subsection{Relaxation time treatment of the spin-3/2 BC model with quenched crystal field}\label{results1}

\begin{figure*}
\includegraphics[width=13.0cm,height=11.0cm]{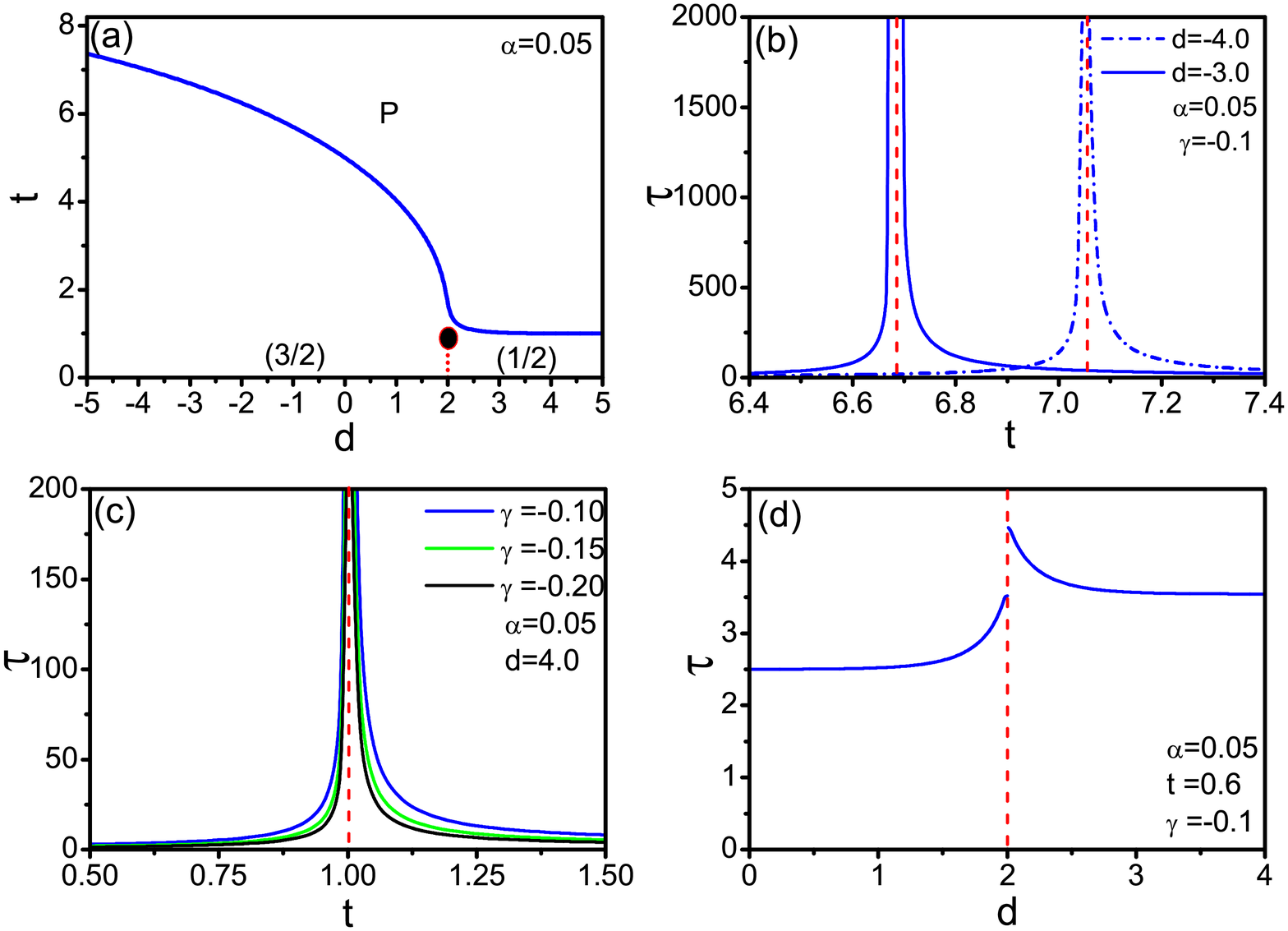}
\caption{(a) Phase diagram of the spin-3/2 BC model with quenched random crystal field for $\alpha=0.05$. The full and dotted lines denote the continuous and discontinuous phase  transitions,  respectively. The solid circle represents the isolated critical point. (b) Thermal variation of the relaxation time for different values of the crystal fields near the critical point. (c) Effects of the $\gamma$ parameters on the relaxation time in the neighborhood of critical point. (d) Crystal field variation of the relaxation time in the neighborhood of discontinuous phase transition.}\label{fig1}
\end{figure*}

Fig. \ref{fig1}(a) represents the phase diagram in the reduced crystal field and reduced
temperature $(d,t)$ plane for $\alpha=0.05$. We should note that this type of phase diagram can
be observed by selecting any value in the range of $0\leq \alpha <1/4$. Here, the full and dotted
lines denote the continuous and discontinuous phase transition points, respectively, and also
the solid circle indicates the isolated critical point. One can clearly see from the Fig. \ref{fig1}(a)
that the ferromagnetic phase separates the paramagnetic phase with a phase boundary line containing
continuous phase transition points, and also phase transition line between
ferromagnetic  $F_{3/2}$ and the ferromagnetic  $F_{1/2}$ phases consists of only first order
phase transition points at the relatively lower temperatures. Fig. \ref{fig1}(b)
shows the thermal variation of the relaxation time $\tau$ in the neighborhood of order-disorder phase
transition temperatures for  selected values of $d$, $\gamma$ and $\alpha$. In this figure, the solid and
dashed dotted curves correspond to values of $d=-3.0$  and $-4.0$ for $\alpha=0.05$ and
$\gamma=-0.1$, respectively.  The vertical dashed lines refer to the  phase transition temperatures for each value
of the reduced crystal field. As seen in Fig. \ref{fig1}(b) $\tau$ grows rapidly with increasing
temperature and diverges as the temperature approaches the order-disorder phase-transition  points for
both crystal field values $d=-4.0$ and $-3.0$. In order to understand the influences of the
Onsager rate coefficient on the relaxation time, we give the Fig. \ref{fig1}(c) for varying $\gamma$ values,
and selected values of $\alpha=0.05$ and $d=4.0$. On both sides of the critical temperature,
illustrated by a vertical dashed line, the relaxation time is divergent for all $\gamma$ values.
It is worthwhile to stress that even though the divergence of $\tau$ gets pushed away from the critical
temperature when the absolute  value of the Onsager rate coefficient decreases, there exists
a qualitative consistency between their thermal variations. Aforementioned behaviors are in accordance
with the previously published studies \cite{erdem1, erdem2, keskin, canko, gulpinar1, gulpinar2}.
On the other hand, Fig. \ref{fig1}(d) corresponds to crystal field dependence of $\tau$ near the first order phase
transition point between ferromagnetic $F_{3/2}$ and $F_{1/2}$ phases  for values
of  $\alpha=0.05$, $t=0.6$ and $\gamma=-0.1$. In contrary to treatment of the second order phase transition, $\tau$
exhibits a  finite discontinuity at the first order phase transition point.

\begin{figure*}
\includegraphics[width=13.0cm,height=6.0cm]{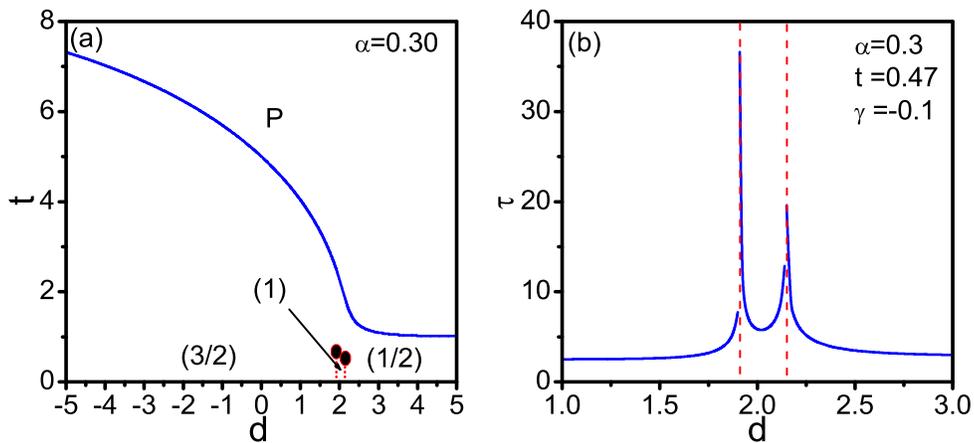}
\caption{(a) Phase diagram of the spin-3/2 BC model with quenched random crystal field for $\alpha=0.30$. The full and dotted lines denote the continuous and discontinuous phase transition points respectively. The solid circles represent the isolated critical points. (b) Crystal field variation of the relaxation time in the neighborhood of discontinuous phase transitions.}\label{fig2}
\end{figure*}

In the following analysis, let us focus on the influences of $\alpha$ parameter on the relaxation dynamics of the
considered magnetic system. In Fig. \ref{fig2}(a), we plot the phase diagram in $(d, t)$ plane
with selected value of $\alpha=0.3$. It can be easily seen from the Fig. \ref{fig2}(a) that there exists
two successive first order phase transitions at the relatively low temperature.   The first transition is from
ferromagnetic $F_{3/2}$ phase to ferromagnetic $F_{1}$ phase while the second transition is between
ferromagnetic $F_{1}$ and  $F_{1/2}$ phases. As mentioned in section \ref{formulation}, similar type of phase
diagrams can be obtained by focusing the range of $1/4\leq \alpha <1$. On the other side, Fig. \ref{fig2}(b)
demonstrates that the $\tau$ exhibits two successive infinite discontinuous for values
of $\alpha=0.3$, $t=0.47$ and $\gamma=-0.1$ in which the first one is between ferromagnetic $F_{3/2}$ and
ferromagnetic $F_{1}$  phases and second  one is between ferromagnetic $F_{1}$ and ferromagnetic $F_{1/2}$.

\begin{figure*}
\includegraphics[width=13.0cm,height=11.0cm]{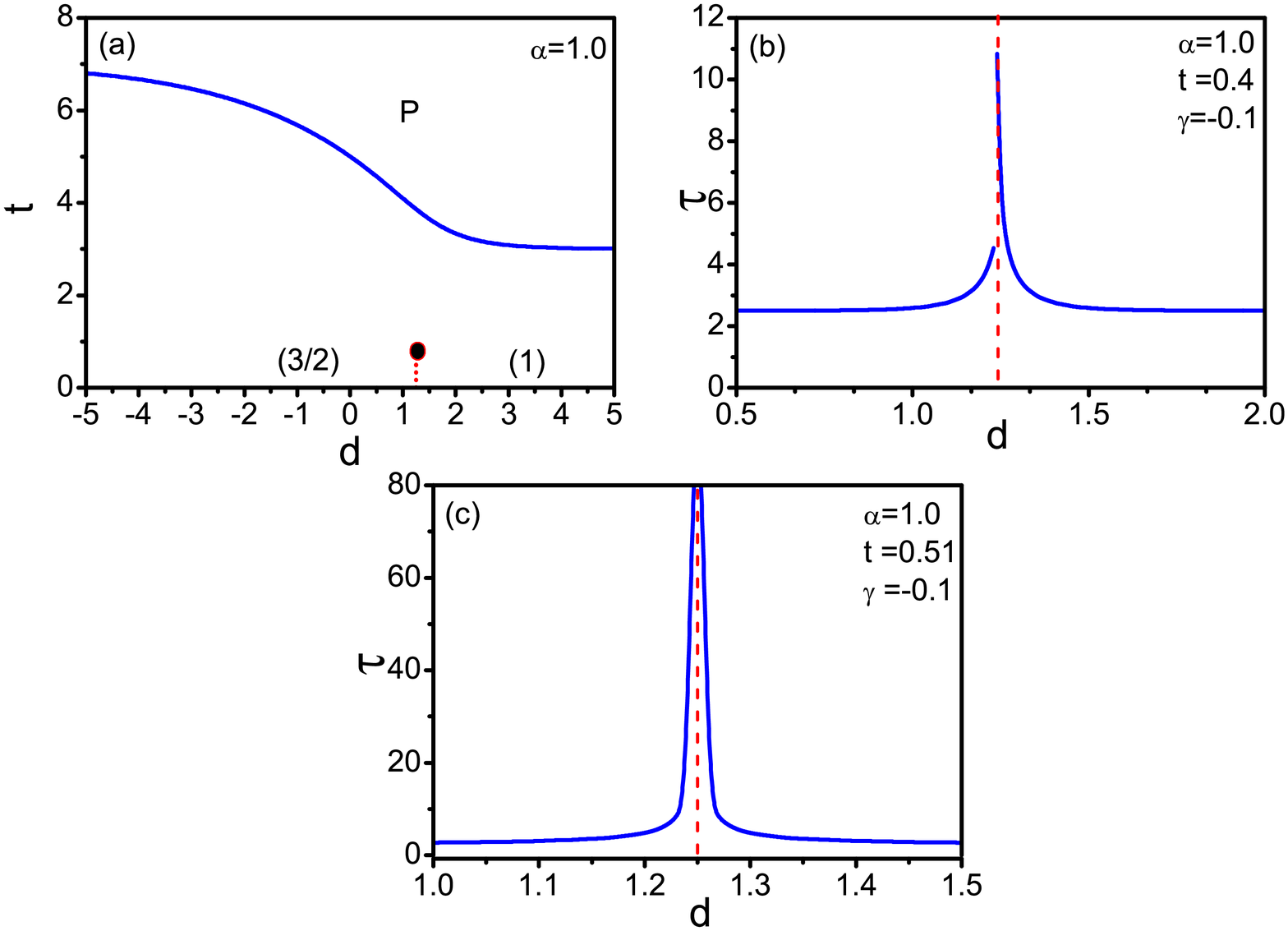}
\caption{(a) Phase diagram of the spin-3/2 BC model with quenched random crystal field for $\alpha=1.0$. The full and dotted lines denote the continuous and discontinuous phase transition points respectively. The solid circle represents the isolated critical point. Crystal field variations of the relaxation time in the vicinity of  discontinuous  (b) and isolated critical point (c).}\label{fig3}
\end{figure*}

In Fig. \ref{fig3}(a), for a bigger $\alpha$ parameter, for example $\alpha=1.0$, a discontinuous phase transition
line exists separating $F_{3/2}$ and $F_{1}$ phases. The crystal field variation of the $\tau$  is shown in Fig.
\ref{fig3}(b) corresponding to the same $\alpha$ parameter for values of $t=0.4$ and $\gamma=-0.1$.
It can be clearly seen from the Fig. \ref{fig3}(b) that as the crystal field  increases, $\tau$ makes a sharp cusp
at the critical crystal field value. As another characteristic feature of the considered system we have examined
the crystal field variation of $\tau$ in the neighborhood of isolated critical point. As it was expected, in Fig. \ref{fig3}(c),
we found that $\tau$ shows a divergence treatment for selected values of parameters such as $\alpha=1.0$, $t=0.51$ and $\gamma=-0.1$.

\begin{figure*}
\includegraphics[width=13.0cm,height=8.0cm]{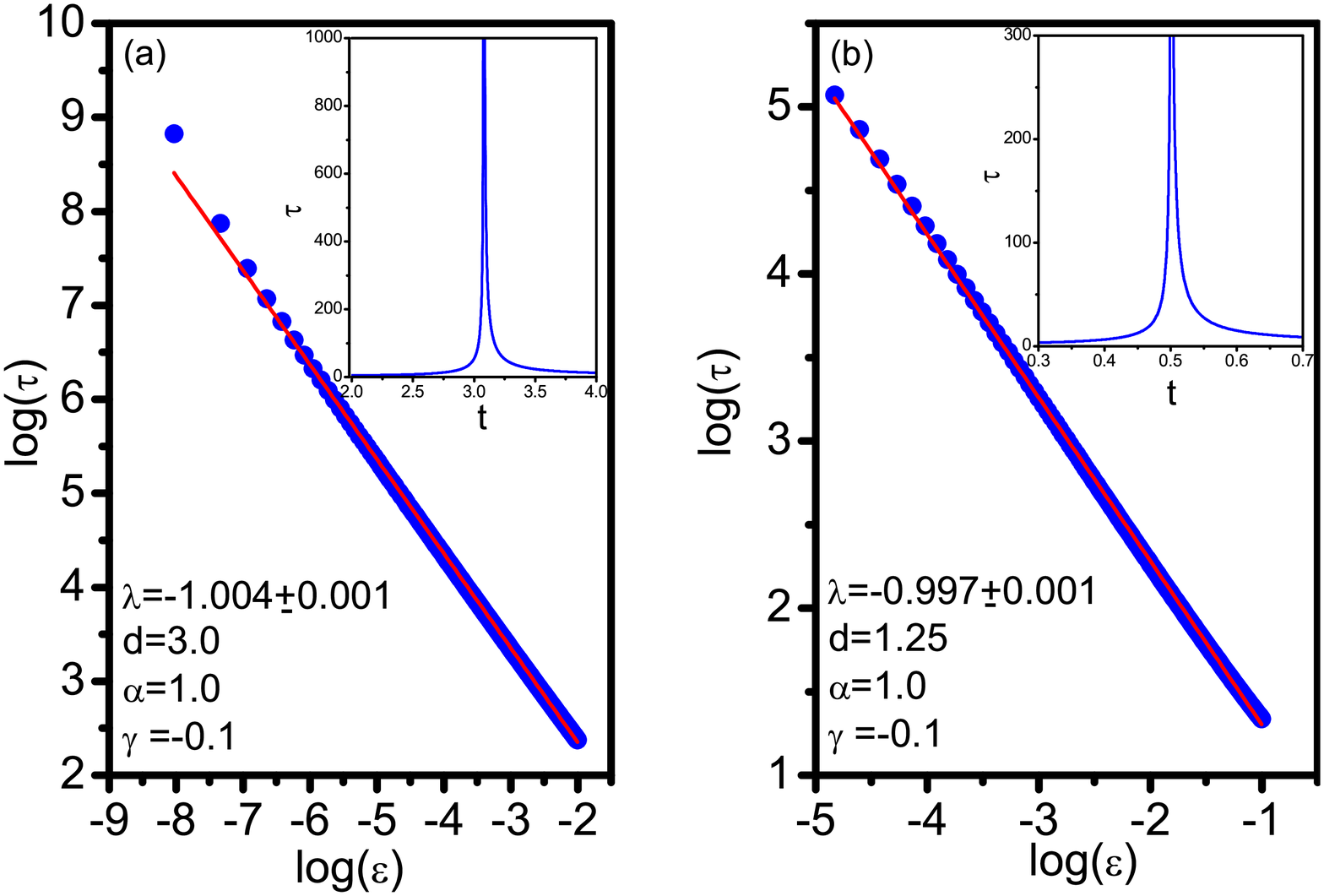}
\caption{Log-log plot of  relaxation time vs. reduced parameter, and in the inset, relaxation time vs. temperature are shown near the continuous phase transition point (a) and near the isolated critical  point (b).}\label{fig4}
\end{figure*}

The divergence of the relaxation time  on both sides of the critical point can be described by the critical exponents.
Thus, we may assume that, for temperatures lower than the transition temperatures,  $\tau$ obeys the  law of the form
$\tau(t)\approx (t_{c}-t)^{\lambda}$ and values of the $\lambda$ can be obtained from the plot of $\log(\tau)-\log(\epsilon)$,
where $\epsilon=\frac{t_{c}-t}{t_{c}}$ \cite{reichl}. In Fig. \ref{fig4}(a), we have plotted $\tau$ versus dimensionless
parameter $(\epsilon)$ in the neighborhood of continuous phase transition point, and we have obtained only one linear part
on this  plot for $\tau$. It is found that the slope of the curve equals to $-1.004\pm0.001$ indicating that $\tau$ diverges
with a mean field exponent of  $-1.004\pm0.001$ as $(t_{c}-t)^{-1.004\pm0.001}$ for the selected values of parameters.
In addition to these,  a similar analyze has been done for isolated critical point in Fig. \ref{fig4}(b). It is
deduced from the log-log plot of $\tau$ versus dimensionless  parameter  that the slope of the curve equals to $-0.997\pm0.001$ which
indicates that the $\tau$ of the spin-3/2 BC model with quenched random crystal field diverges with a mean-field exponent
of $-0.997\pm0.001$ as $(t_{c}-t)^{-0.997\pm0.001}$ for values of $d=1.25$, $\alpha=1.0$ and $\gamma=-1.0$.

\begin{figure*}
\includegraphics[width=13.0cm,height=11.0cm]{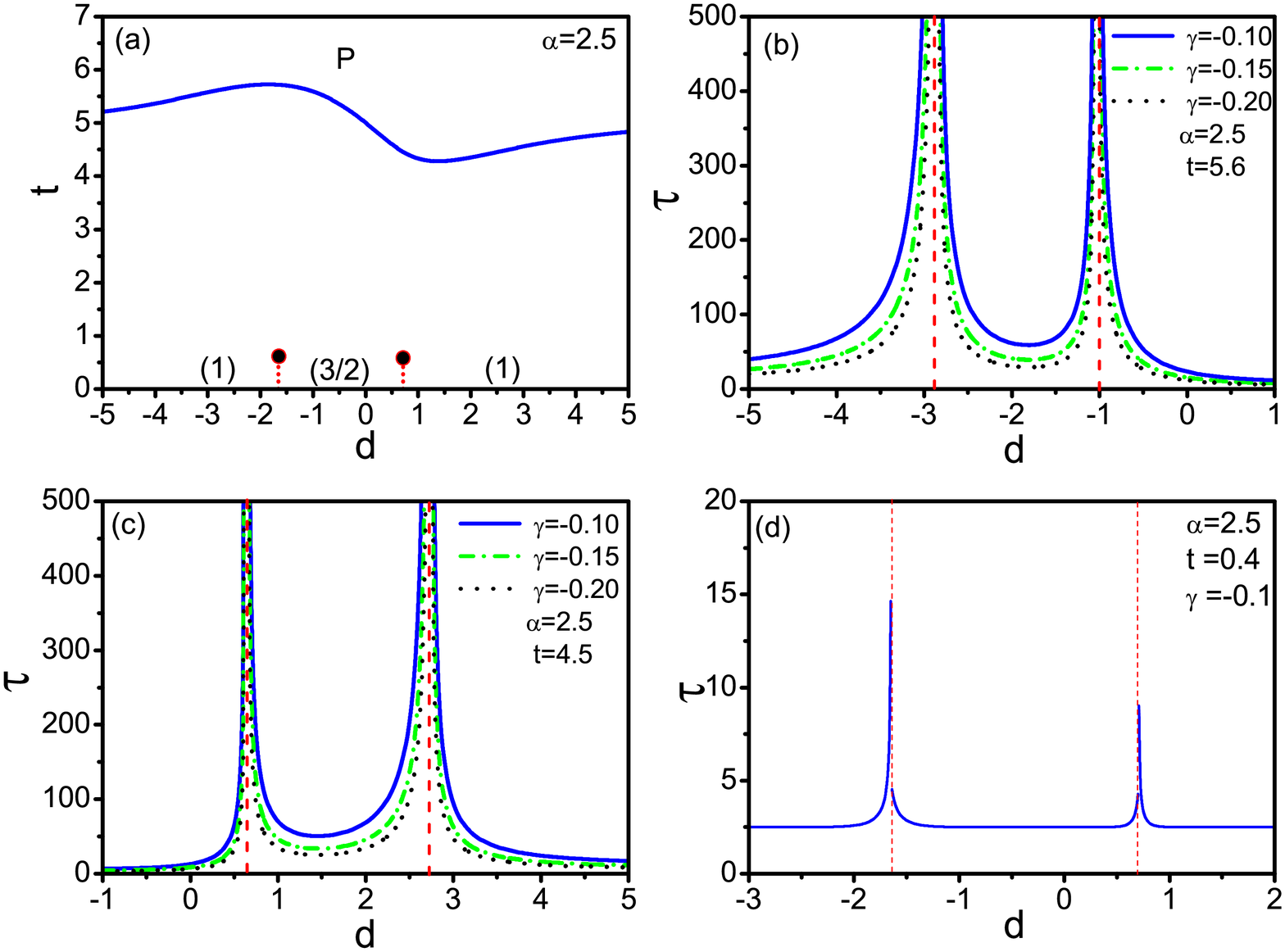}
\caption{(a) Phase diagram of the spin-3/2 BC model with quenched random crystal field for $\alpha=2.5$. The full and dotted lines denote the continuous and discontinuous phase  transition points, respectively. The solid circles represent the isolated critical points.  (b) Effect of the kinetic rate coefficient on the reentrant behavior at $t=5.6$ (c) Effect of the kinetic rate coefficient on the reentrant behavior at $t=4.5$ (d) Crystal field variation of the relaxation time in the neighborhood discontinuous phase transitions.}\label{fig5}
\end{figure*}

It can be seen clearly from the Fig. \ref{fig5}(a) that the magnetic system exhibits reentrant behavior at the
relatively higher temperature regions for $\alpha=2.5$.  This behavior originates from the competition between
negative and positive values of crystal field interactions. And also, two discontinuous phase transition lines
occur at  the low temperature regions.  The first one  is  between  the ferromagnetic $F_{1}$ and
ferromagnetic $F_{3/2}$ phases, and the second one  is from the  ferromagnetic $F_{3/2}$ to $F_{1}$ phases.
In order to show the reentrant behavior on the relaxation times,  we give the Fig. \ref{fig5}(b) for varying
values of $\gamma$ and  selected values of  $t$, and $d$. As was expected, two successive divergences
indicating two successive  second order phase transition  points exist  in  Fig. \ref{fig5}(b) for  values
of $\alpha=2.5$, $t=4.5$ where  the first one is from ordered phase to disordered phase
while the other is between the disordered and ordered  phases,  respectively. It is necessary to say that as
the absolute  value of the Onsager rate coefficient  decreases $(\gamma)$, divergence of $\tau$ gets pushed
away from the critical temperature. As a final investigation, we focused on the low temperature
region and we gave the  crystal field variation of $\tau$ in Fig. \ref{fig5}(c). One can see from
the figure Fig. \ref{fig5}(d), $\tau$ shows two infinite discontinuous corresponding to phase transition points, as
crystal field interaction increases for selected values such as $\alpha=2.5$, $t=0.4$ and $\gamma=-0.1$.

\subsection{Complex magnetic susceptibility treatment of the spin-3/2 BC model with quenched crystal field}\label{results2}
\begin{figure*}
\includegraphics[width=16.0cm,height=11.0cm]{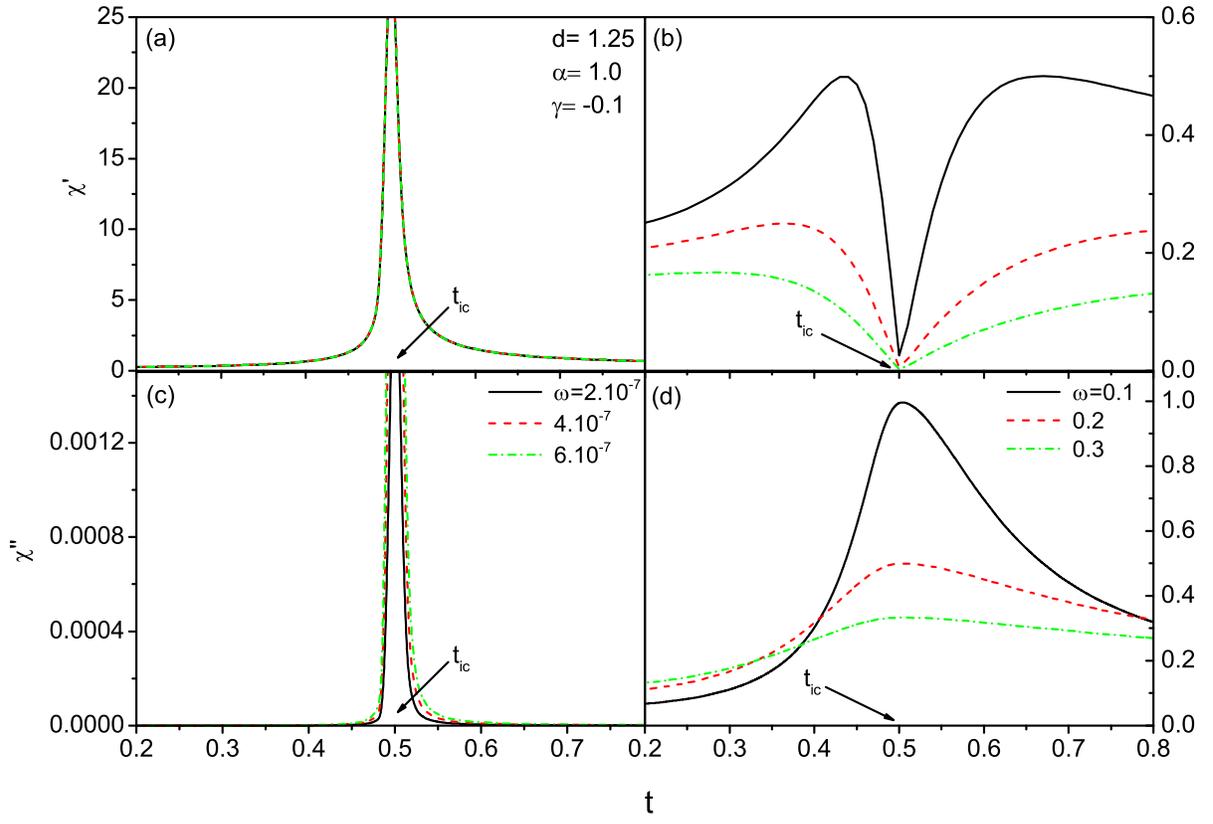}
\caption{Temperature variations of magnetic dispersion and absorption factors for the low (a and c) and high frequency (b and d) region
at several values of the frequency $(\omega)$ in the vicinity of isolated critical point while $d=1.25, \alpha=1.0$ and $\gamma=-0.1$.
The arrow refers to the isolated critical point.}\label{fig6}
\end{figure*}

Figs. \ref{fig6}(a-d) illustrate the real and imaginary parts of the ac susceptibility in the vicinity of
isolated critical point for both low and high frequency regions and
 values of $d=1.25$, $\alpha=1.0$  and $\gamma=-0.1$. In Figs. \ref{fig6}(a-b) the
temperature variations of the magnetic dispersion factors are given, and, one can easily see from the figures  that
the $\chi^{'}$ increases rapidly with increasing temperature and exhibits a divergence near the isolated critical point
for the low frequency values. On the other hand, the $\chi^{'}$ shows a shallow dip at the isolated critical point
for the relatively higher applied field frequencies. In addition, two frequency dependent local maxima
occur at  before and after the isolated critical point, respectively (see Fig. \ref{fig6}(b)). From the
Figs. \ref{fig6}(c-d) we see that the $\chi^{''}$ exhibits a divergence at the isolated critical point for the low frequency
regions, while it displays a local maxima for the relatively higher applied field frequencies. By comparing
figures \ref{fig6}(a) and (c) one can see that the divergence treatment of $\chi^{'}$ is independent of the applied field frequency
whereas divergence of $\chi^{''}$ sensitively depends on the frequency.  On the other hand,
it is possible to say that  the similar ac magnetic susceptibility behavior can be observed
by focusing any second order phase transition point.

\begin{figure*}
\includegraphics[width=16.0cm,height=11.0cm]{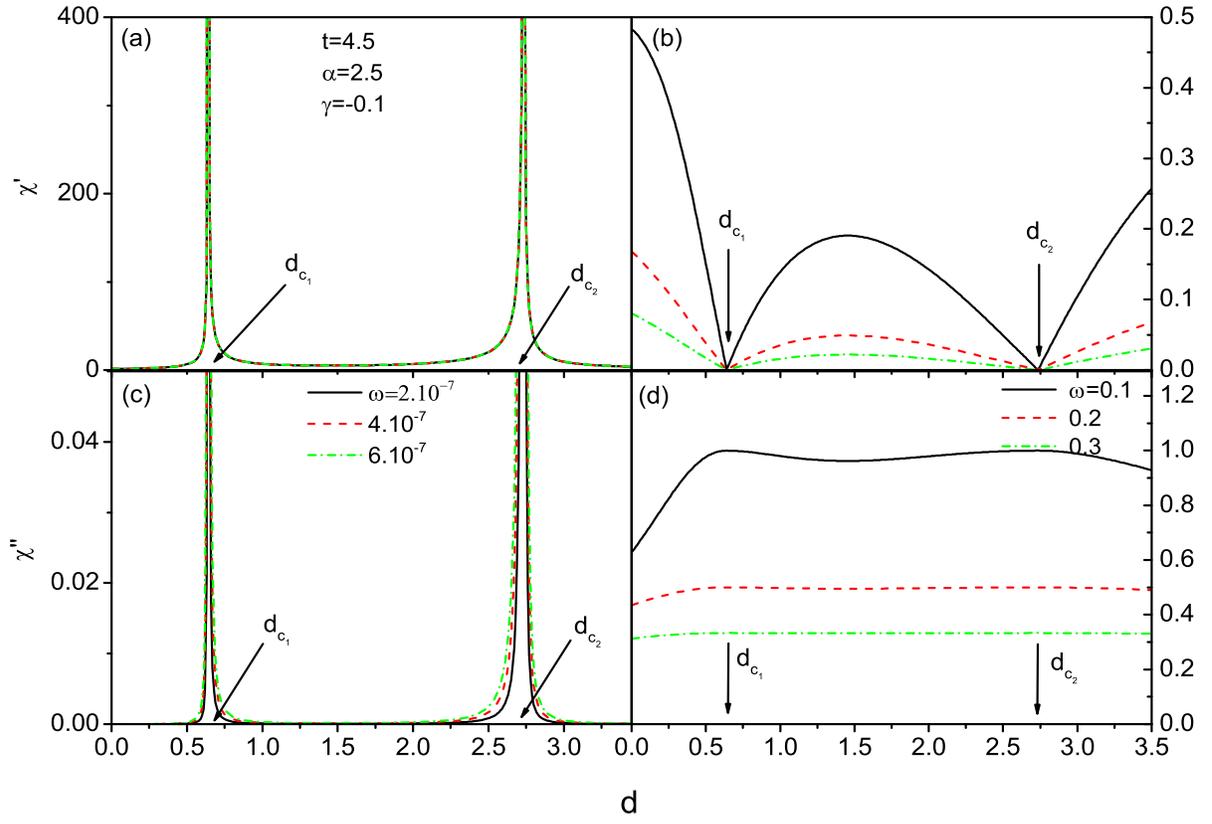}
\caption{Crystal field variations of magnetic dispersion and absorption factors for the low (a and c) and high frequency (b and d) regions
at several values of the frequency $(\omega)$ in the vicinity of reentrant region while $t=4.5, \alpha=2.5$ and $\gamma=-0.1$.
The arrows refer to the critical crystal field values.}\label{fig7}
\end{figure*}

In the following analysis, let us investigate the reentrant behavior properties on the magnetic dispersion and
absorption factors corresponding to the phase diagram in Fig. \ref{fig5}(a). In Fig. \ref{fig7}(a) we represent the
crystal field variation of the magnetic absorption factor for selected values of $t, \gamma$ and $\alpha$ with
different applied field frequencies in the low frequency region. One can observe from the figure that the magnetic system
exhibits two successive divergences behavior indicating two successive  second order phase transition  points where
the first one is from ordered phase to disordered phase while the other is between the disordered and ordered
phases,  respectively. On the other hand, as seen from figure \ref{fig7}(b), the magnetic absorption factor has two
frequency dependent shallow dips  at the second order phase transition points in the high frequency region. In addition,
it is useful to mention that as we increase the applied field frequency, the maximum located at a finite crystal field
value in the ordered and disordered phases decreases.

\begin{figure*}
\includegraphics[width=10.0cm,height=9.0cm]{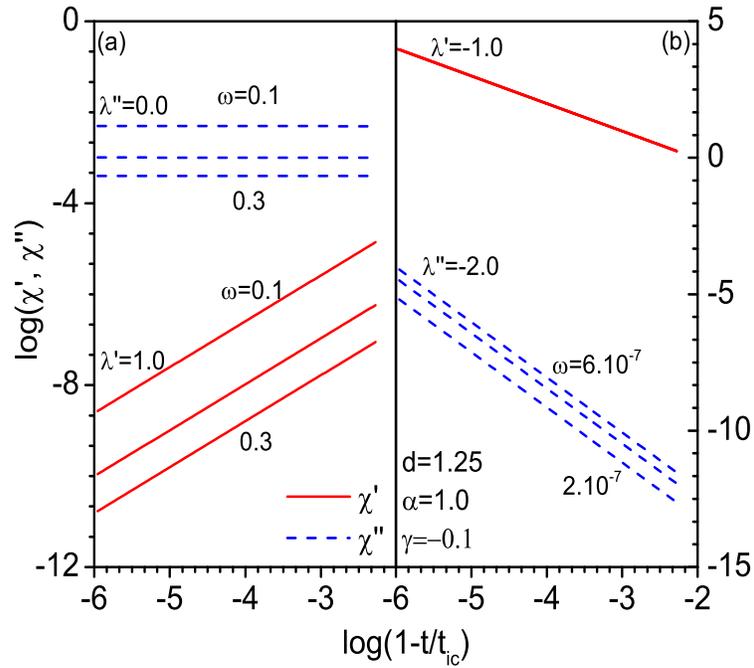}
\caption{Double logarithmic (log-log) plots of magnetic dispersion and absorption factors versus $\mathrm{(1-t/t_{ic})}$ at several values in the high (a)  and low (b)
frequency regions while $d=1.25, \alpha=1.0$ and $\gamma=-0.1$. Solid and dashed lines denote the $\log(\chi')$ and $\log(\chi'')$, respectively.
}\label{fig8}
\end{figure*}

As a final investigation, dynamical mean field critical exponents for isolated critical point are calculated for
magnetic dispersion  and absorption factors for both low and high frequency regions in Fig. \ref{fig8}. It is known that
the critical behaviors of $\mathrm{\chi'}$ and $\mathrm{\chi''}$ exponents in the neighborhood of critical point as well as isolated
critical point are characterized by the critical exponents. Thus, we may assume that, for example, when $\mathrm{t\rightarrow t_{ic}}$,
$\mathrm{\chi'}$ and $\mathrm{\chi''}$  follow scaling laws of the form $\mathrm{\chi'\sim |t-t_{ic}|^{\lambda'}}$ and $\mathrm{\chi''\sim |t-t_{ic}|^{\lambda''}}$,
respectively. Here, $\lambda'$,  $\lambda''$ and $\mathrm{t_{ic}}$ are dynamical mean field critical exponents of the magnetic  dispersion and absorption
factors and isolated critical point of the considered system, respectively. In order to calculate the value of the dynamical critical exponents one should sketch double logarithmic
plot of the related quantity versus $\mathrm{log(1-t/t_{ic})}$ and find the slope of the linear part of this curve. This definition is
valid for all values of the exponents where the negative values correspond to the divergence of the related quantity as $\mathrm{1-t/t_{ic}}$  goes
to zero and the positive values correspond to logarithmic divergence, cusps or jump singularities \cite{erdem1, reichl}. Fig. \ref{fig8}(a) represents the
the plots of $\mathrm{log(\chi')}$ and $\mathrm{log(\chi'')}$ versus $\mathrm{log(1-t/t_{ic})}$ at several values of applied field
frequency corresponding  to high region in the vicinity of isolated critical point for  values of  $\alpha=1.0$, $d=1.25$ and $\gamma=-0.1$. We should mention here that we have found only one linear part on each of these  $\mathrm{\log-\log}$ plots, namely, one value for $\mathrm{\lambda'}$, and one
value for $\mathrm{\lambda''}$, respectively. It is clearly seen from the figure that the behavior
of $\mathrm{log(\chi')}$ and $\mathrm{log(\chi'')}$ are dependent the applied field frequency, and the slopes of the $\mathrm{log(\chi')}$ and $\mathrm{log(\chi'')}$ are equal to $1.0$ and $0$, respectively. Namely, although the behaviors of
the magnetic  dispersion and absorption  factors depend on the field frequency in the high frequency region, their slopes
are not dependent the $\omega$. This scaling form $(\lambda'=1.0)$ verifies the convergence of the  $\mathrm{\chi'}$ to
zero in the high-frequency  region in the neighborhood of isolated critical point
(see Fig. \ref{fig6}(b)). In addition, scaling form $(\lambda''=0.0)$ of the $\mathrm{\chi''}$  implies
the  existence a local maxima in the high frequency region, in other words, it is in accordance with Fig. \ref{fig6} (d).
On the other hand, in Fig. \ref{fig8} (b), we depict the the plots of $\mathrm{log(\chi')}$ and $\mathrm{log(\chi'')}$
versus $\mathrm{log(1-t/t_{ic})}$ at several values of applied field frequency corresponding  to low region
in the vicinity of isolated critical point for  the same values with Fig. \ref{fig8} (a). One can figure out
from Fig.\ref{fig8} (b) that the logarithmic behavior of magnetic dispersion factor does not depend on
the applied field frequency  whereas the magnetic absorption factor changes with $\omega$ in the low frequency
region. The mean field dynamic critical exponent of the $\mathrm{log(\chi')}$
and $\mathrm{log(\chi'')}$ are equal to $-1$ and $-2$, respectively. It is obvious that these scaling
forms $(\mathrm{\lambda'=-1.0, \lambda''=-2.0})$  correspond to divergence behavior in the vicinity of isolated
critical point in the low frequency region and for  selected system parameters.

\section{Concluding Remarks}\label{conclude}
In this study, we have studied the nonequilibrium dynamics of the spin-3/2 BC model with quenched random crystal field
by means of Onsager irreversible thermodynamics. In order to obtain the Gibbs free energy of the considered disordered system,
we used the MFA, and by benefiting from the minimization condition of Gibbs free energy, we found the mean field
equation of state and we represented the  phase diagram in the $(d, t)$ plane for  different $\alpha$ parameters.
After that, we have calculated the Gibbs free-energy production in the irreversible process and by means of this generalized
force and flux are defined in irreversible thermodynamics limit. Then, the kinetic  equation for  the magnetization is
obtained by using linear  response theory. From the solution of the kinetic equation in the vicinity of equilibrium states,
a relaxation time describing the nonequilibrium dynamics of the considered system is obtained. In order to understand the
critical behavior of the relaxation phenomena, the temperature and also crystal field dependencies of the relaxation time are determined
in the vicinity of the phase-transition points. We found that $\tau$ grows rapidly with increasing temperature and
diverges as the temperature approaches the order-disorder phase-transition points as well as in the vicinity of isolated critical
point for selected Hamiltonian parameters and Onsager rate coefficients. And also we calculated the dynamical mean field critical
exponents for order-disorder phase transition and isolated critical points, respectively and it is found that the slopes of the curves are
equal to $-1.004\pm0.001$  and $-0.997\pm0.001$  indicating that $\tau$ diverges with increasing temperature. In contrary to
treatment of the second order phase transition, $\tau$ exhibits a finite discontinuity at the first order phase transition
point for fixed temperature value. In addition, as mentioned above we have touched upon the influences of Onsager rate coefficient
on the $\tau$ behavior, and it is observed that even though divergence of $\tau$ gets pushed away from the critical temperature
when the absolute  value of the Onsager rate coefficient decreases $(\gamma)$, there exists a qualitative consistency between
their thermal variations.  Aforementioned behaviors are in accordance with the previously published studies
\cite{erdem1, erdem2, keskin,  canko, gulpinar1, gulpinar2}.

Moreover, we have analyzed the complex magnetic susceptibility treatment of the  spin-3/2 BC model with quenched
random crystal field under the time dependent oscillating magnetic field. It is worthwhile to mention that the amplitude of
applied magnetic field is so small that we made use of linear response theory in studying the magnetic relaxation process.
We especially have touched upon the thermal variation of complex magnetic susceptibility near the isolated critical point
for varying applied field frequencies. We found that the magnetic dispersion and absorption factors grow drastically and
diverges with increasing temperature in the low frequency regions (see  Figs. \ref{fig6}(a) and (c)). In addition, we can say that
the divergence treatment of $\chi^{'}$ is independent of the  field frequency whereas divergence of $\chi^{''}$ sensitively
depends on the frequency.  On the other hand, the $\chi^{'}$ and $\chi^{'}$ show  a shallow dip and a local maxima at the isolated
critical point for the relatively higher applied field frequencies, respectively. It is also observed that
two frequency dependent local maxima occur at  before and after the isolated critical point, respectively.
Moreover, we have investigated the crystal field dependence of the complex magnetic susceptibility
in the neighborhood of reentrant region for selected Hamiltonian parameters
and we found that the magnetic system exhibits two successive divergences behavior indicating two successive
second order phase transition points in the low frequency region. For higher frequency region,
the magnetic absorption factor has two frequency dependent shallow dips at the second order phase transition points.
In addition, it is useful to mention that as we increase the applied field frequency, the maximum located at a finite
crystal field value in the ordered and disordered phases decreases. As a
final investigation  we have calculated the dynamical mean field critical
exponents of complex magnetic susceptibility for isolated critical point.
We also mention that a qualitatively good agreement exists between our findings and previously published
studies \cite{barry2, erdem1, gulpinar3}.

\section*{Acknowledgements}
The numerical calculations reported in this paper were performed at T\"{U}B\.{I}TAK ULAKBIM (Turkish agency),
High Performance and Grid Computing Center (TRUBA Resources).

\section{Appendix}
The coefficients $\Lambda_{i} (i=1,..,4)$ can be easily calculated by using Taylor expansion
and  are given as follows:

\begin{equation*}
\begin{array}{lcl}
\Lambda_{1}=\displaystyle\left(\frac{\partial^2g}{\partial m^2}\right)_{eq}=\\
\\
\displaystyle z+\displaystyle\frac{t}{2}\left(\frac{\frac{3\exp\left(\frac{-9\ell_{1}}{4}\right)\ell_{5}z+
\exp\left(\frac{-\ell_{1}}{4}\right)\ell_{6}z}{t}}{2\exp\left(\frac{-9\ell_{1}}{4}\right)\ell_{3}+2\exp\left(\frac{-\ell_{1}}{4}\right)\ell_{4}}\right)^2\\
\\\displaystyle +\frac{t}{2}\left(\frac{\frac{3\exp\left(\frac{-9\ell_{2}}{4}\right)\ell_{5}z+
\exp\left(\frac{-\ell_{2}}{4}\right)\ell_{6}z}{t}}{2\exp\left(\frac{-9\ell_{2}}{4}\right)\ell_{3}+2\exp\left(\frac{-\ell_{2}}{4}\right)\ell_{4}}\right)^2\\
\\
\displaystyle -\frac{t}{2}\left(\frac{\frac{9\exp\left(\frac{-9\ell_{1}}{4}\right)\ell_{3}z^2+
\exp\left(\frac{-\ell_{1}}{4}\right)\ell_{4}z^2}{2t^2}}{2\exp\left(\frac{-9\ell_{1}}{4}\right)\ell_{3}+2\exp\left(\frac{-\ell_{1}}{4}\right)\ell_{4}}\right)\\
\\
\displaystyle -\frac{t}{2}\left(\frac{\frac{9\exp\left(\frac{-9\ell_{2}}{4}\right)\ell_{3}z^2+
\exp\left(\frac{-\ell_{2}}{4}\right)\ell_{4}z^2}{2t^2}}{2\exp\left(\frac{-9\ell_{2}}{4}\right)\ell_{3}+2\exp\left(\frac{-\ell_{2}}{4}\right)\ell_{4}}\right),
\\
\\
\displaystyle\Lambda_{2}=\left(\frac{\partial g}{\partial b}\right)_{eq}=-m,
\\
\\
\displaystyle\Lambda_{3}=\left(\frac{\partial^2g}{\partial b^2}\right)_{eq}=0,
\\
\\
\displaystyle\Lambda_{4}=\left(\frac{\partial^2g}{\partial m \partial b}\right)_{eq}=-1.

\end{array}
\end{equation*}

\newpage
\clearpage

\end{document}